\documentclass[runningheads]{llncs}
\usepackage{latexsym}
\usepackage{amsfonts}
\usepackage{amsmath}
\usepackage{amssymb}
\usepackage{graphicx}
\usepackage{float}
\usepackage{epsfig}
\usepackage{subfigure}

\def\reals{\mathbb{R}}
\def\rationals{\mathbb{Q}}

\newcommand{\T}{\mathbf{T}}
\newcommand{\s}{\mathbf{s}}
\renewcommand{\t}{\mathbf{t}}

\newcommand{\pbr}{\mathop{\rm pbr}\nolimits}
\newcommand{\PBR}{{\cal B}}

\newcommand{\poly}{\mathop{\rm poly}\nolimits}
\newcommand{\argmax}{\mathop{\rm argmax}\nolimits}

\begin{document}
\title{
Computing Good Nash Equilibria in Graphical Games
\thanks {Supported by the
EPSRC research grants ``Algorithmics of Network-sharing Games''
and
``Discontinuous Behaviour in the Complexity of randomized
Algorithms''.
}}

\author{Edith Elkind$^{1}$\
\and Leslie Ann Goldberg$^{2}$ 
\and Paul W.~Goldberg$^{2}$}

\institute{
$^{1}$Hebrew University of Jerusalem, Israel, and
University of Southampton, Southampton, SO17 1BJ, U.K.
\\
$^{2}$ Department of Computer Science, University of 
Liverpool\\
           Liverpool L69 3BX, United Kingdom}

\maketitle
\begin{abstract}
This paper addresses the problem of fair equilibrium selection in
graphical games. Our approach is based on the data structure called
the {\em best response policy}, which was proposed by Kearns et
al.~\cite{kls} as a way to represent all Nash equilibria of a
graphical game. In~\cite{egg}, it was shown that the best response
policy has polynomial size as long as the underlying graph is a path.
In this paper, we show that if the underlying graph is a
bounded-degree tree and the best response policy has polynomial size
then there is an efficient algorithm which constructs a Nash
equilibrium that guarantees certain payoffs to all
participants. Another attractive solution concept is a Nash
equilibrium that maximizes the social welfare.  We show that, while
exactly computing the latter is infeasible (we prove that solving this
problem may involve algebraic numbers of an arbitrarily high degree),
there exists an FPTAS for finding such an equilibrium as long as the
best response policy has polynomial size. These two algorithms can be
combined to produce Nash equilibria that satisfy various fairness
criteria.
\end{abstract}

\section{Introduction}

In a large community of agents, an agent's behavior is not likely to
have a direct effect on most other agents: rather, it is just the
agents who are close enough to him that will be affected. However, as
these agents respond by adapting their behavior, more agents will feel
the consequences and eventually the choices made by a single agent
will propagate throughout the entire community.

This is the intuition behind {\em graphical games}, which were
introduced by Kearns, Littman and Singh in~\cite{kls} as a compact
representation scheme for games with many players. In an $n$-player
graphical game, each player is associated with a vertex of an
underlying graph $G$, and the payoffs of each player depend on his
action as well as on the actions of his neighbors in the graph.  If
the maximum degree of $G$ is $\Delta$, and each player has two actions
available to him, then the game can be represented using
$n2^{\Delta+1}$ numbers. In contrast, we need $n2^n$ numbers to
represent a general $n$-player 2-action game, which is only practical
for small values of $n$.  For graphical games with constant $\Delta$,
the size of the game is linear in~$n$.

One of the most natural problems for a graphical game is that of
finding a Nash equilibrium, the existence of which follows from Nash's
celebrated theorem (as graphical games are just a special case of
$n$-player games). The first attempt to tackle this problem was made
in~\cite{kls}, where the authors consider graphical games with two
actions per player in which the underlying graph is a bounded-degree
tree. They propose a generic algorithm for finding Nash equilibria
that can be specialized in two ways: an exponential-time algorithm for
finding an (exact) Nash equilibrium, and a w
fully polynomial time
approximation scheme (FPTAS) for finding an approximation to a Nash
equilibrium.  For any $\epsilon>0$ this algorithm outputs an {\it
$\epsilon$-Nash equilibrium,} which is a strategy profile in which no
player can improve his payoff by more than $\epsilon$ by unilaterally
changing his strategy.

While $\epsilon$-Nash equilibria are often easier to compute than
exact Nash equilibria, this solution concept has several drawbacks.
First, the players may be sensitive to a small loss in payoffs, so the
strategy profile that is an $\epsilon$-Nash equilibrium will not be
stable. This will be the case even if there is only a small subset of
players who are extremely price-sensitive, and for a large population
of players it may be difficult to choose a value of $\epsilon$ that
will satisfy everyone.  Second, the strategy profiles that are close
to being Nash equilibria may be much better with respect to the
properties under consideration than exact Nash equilibria. Therefore,
the (approximation to the) value of the best solution that corresponds
to an $\epsilon$-Nash equilibrium may not be indicative of what can be
achieved under an exact Nash equilibrium.  This is especially
important if the purpose of the approximate solution is to provide a
good benchmark for a system of selfish agents, as the benchmark
implied by an $\epsilon$-Nash equilibrium may be unrealistic.  For
these reasons, in this paper we focus on the problem of computing
exact Nash equilibria.

Building on ideas of~\cite{lks}, Elkind et al.~\cite{egg} showed how
to find an (exact) Nash equilibrium in polynomial time when the
underlying graph has degree~2 (that is, when the graph is a collection
of paths and cycles). By contrast, finding a Nash equilibrium in a
general degree-bounded graph appears to be computationally
intractable: it has been shown (see \cite{cd,gp,dgp}) to be complete
for the complexity class PPAD. \cite{egg} extends this hardness
result to the case in which the underlying graph has bounded
pathwidth.

A graphical game may not have a unique Nash equilibrium, indeed it may
have exponentially many. Moreover, some Nash equilibria are more
desirable than others.  Rather than having an algorithm which merely
finds \emph{some} Nash equilibrium, we would like to have algorithms
for finding Nash equilibria with various socially-desirable
properties, such as maximizing overall payoff or distributing profit
fairly.

A useful property of the data structure of~\cite{kls} is that it
simultaneously represents the set of {\em all} Nash equilibria of the
underlying game. If this representation has polynomial size (as is the
case for paths, as shown in~\cite{egg}), one may hope to extract from
it a Nash equilibrium with the desired properties. In fact,
in~\cite{kls} the authors mention that this is indeed possible if one
is interested in finding an (approximate) $\epsilon$-Nash
equilibrium. The goal of this paper is to extend this to exact Nash 
equilibria.

\subsection{Our Results}

In this paper, we study $n$-player $2$-action graphical games on
bounded-degree trees for which the data structure of~\cite{kls} has
size $\poly(n)$. We focus on the problem of finding exact Nash
equilibria with certain socially-desirable properties.  In particular,
we show how to find a Nash equilibrium that (nearly) maximizes the
social welfare, i.e., the sum of the players' payoffs, and we show how
to find a Nash equilibrium that (nearly) satisfies prescribed payoff
bounds for all players.

Graphical games on bounded-degree trees have a simple algebraic
structure.  One attractive feature, which follows from~\cite{kls}, is
that every such game has a Nash equilibrium in which the strategy of
every player is a rational number.  Section~\ref{sec:rational} studies
the algebraic structure of those Nash equilibria that maximize social
welfare.  We show (Theorems~\ref{thm:quad} and~\ref{thm:arbitrary})
that, surprisingly, the set of Nash equilibria that maximize social
welfare is more complex. In fact, for any algebraic number
$\alpha\in[0, 1]$ with degree at most~$n$, we exhibit a graphical game
on a path of length $O(n)$ such that, in the unique social
welfare-maximizing Nash equilibrium of this game, one of the players
plays the mixed strategy~$\alpha$.\footnote{ A related result in a
different context was obtained by Datta~\cite{datta}, who shows that
$n$-player $2$-action games are {\em universal} in the sense that any
real algebraic variety can be represented as the set of totally mixed
Nash equilibria of such games.}
%Footnotes are unpopular, but we really don't want to break the flow with this
%remark, which is not too relevant to the development of our results.
This result shows that it may be difficult to represent an optimal
Nash equilibrium.  
It seems to be a novel feature of the setting we consider here, that
an optimal Nash equilibrium is hard to represent, in a situation
where it is easy to find and represent a Nash equilibrium.

As the social welfare-maximizing Nash equilibrium may be hard to
represent efficiently, we have to settle for an approximation.
However, the crucial difference between our approach and that
of previous papers~\cite{kls,ok,vk} is that we require our algorithm to 
output an
exact Nash equilibrium, though not necessarily the optimal one with
respect to our criteria.  In Section~\ref{bestnash}, we describe an
algorithm that satisfies this requirement. Namely, we propose an
algorithm that for any $\epsilon>0$ finds a Nash equilibrium whose
total payoff is within $\epsilon$ of optimal. It runs in polynomial
time (Theorem~\ref{thm:bestnash},\ref{cor:tree}) for any graphical
game on a bounded-degree tree for which the data structure proposed
by~\cite{kls} (the so-called {\em best response policy}, defined
below) is of size $\poly(n)$ (note that, as shown in~\cite{egg}, this
is always the case when the underlying graph is a path).  More
precisely, the running time of our algorithm is polynomial in $n$,
$P_{\max}$, and $1/\epsilon$, where $P_{\max}$ is the maximum absolute
value of an entry of a payoff matrix, i.e., it is a pseudopolynomial
algorithm, though it is fully polynomial with respect to
$\epsilon$. We show (Section~\ref{multiplicative}) that under some
restrictions on the payoff matrices, the algorithm can be transformed
into a (truly) polynomial-time algorithm that outputs a Nash
equilibrium whose total payoff is within a $1-\epsilon$ factor from
the optimal.

In Section~\ref{bounds}, we consider the problem of finding a Nash
equilibrium in which the expected payoff of each player~$V_i$ exceeds
a prescribed threshold~$T_i$. Using the idea from
Section~\ref{bestnash} we give (Theorem~\ref{thm:besteach}) a fully
polynomial time approximation scheme for this problem.  The running
time of the algorithm is bounded by a polynomial in $n$, $P_{\max}$,
and $\epsilon$.  If the instance has a Nash equilibrium satisfying the
prescribed thresholds then the algorithm constructs a Nash equilibrium
in which the expected payoff of each player $V_i$ is at least
$T_i-\epsilon$.
%I really don't think this is interesting (in fact, I think the section should
% be removed or moved to an appendix)
%%We consider the problem of finding an {\em exact} Nash
%%equilibrium under this constraint, and exhibit an example that
%%shows that this would require irrational numbers to be stored
%%in the data structure.

In Section~\ref{fair}, we introduce other natural criteria for
selecting a ``good'' Nash equilibrium and we show that the algorithms
described in the two previous sections can be used as building blocks
in finding Nash equilibria that satisfy these criteria.  In
particular, in Section~\ref{combining} we show how to find a Nash
equilibrium that approximates the maximum social welfare, while
guaranteeing that each individual payoff is close to a prescribed
threshold.  In Section~\ref{minimax} we show how to find a Nash
equilibrium that (nearly) maximizes the minimum individual payoff.
Finally, in Section~\ref{equalising} we show how to find a Nash
equilibrium in which the individual payoffs of the players are close
to each other.

\subsection{Related Work}

Our approximation scheme (Theorem~\ref{thm:bestnash} and
Theorem~\ref{cor:tree}) shows a contrast between the games that we
study and two-player $n$-action games, for which the corresponding
problems are usually intractable.  For two-player $n$-action games,
the problem of finding Nash equilibria with special properties is
typically NP-hard.  In particular, this is the case for Nash
equilibria that maximize the social welfare~\cite{gz,cs}.  Moreover,
it is likely to be intractable even to approximate such equilibria.
In particular, Chen, Deng and Teng~\cite{cdt} show that there exists
some $\epsilon$, inverse polynomial in $n$, for which computing an
$\epsilon$-Nash equilibrium in 2-player games with $n$ actions per
player is PPAD-complete.

Lipton and Markakis~\cite{lm} study the algebraic properties of Nash
equilibria, and point out that standard quantifier elimination
algorithms can be used to solve them. Note that these algorithms are
not polynomial-time in general.
%%Again, I can't see what the following is for.
%From~\cite{gp}, the solution to
%a degree 3 graphical game can be forced to be a fixpoint of
%a high-degree polynomial. But these are situations where
%hardness results also exist for finding any
%(unrestricted) Nash equilibria.
The games we study in this paper have polynomial-time computable Nash
equilibria in which all mixed strategies are rational numbers, but an
optimal Nash equilibrium may necessarily include mixed strategies with
high algebraic degree.

A {\em correlated equilibrium} (CE) (introduced by
Aumann~\cite{aumann}) is a distribution over vectors of players'
actions with the property that if any player is told his own action
(the value of his own component) from a vector generated by that
distribution, then he cannot increase his expected payoff by changing
his action. Any Nash equilibrium is a CE but the converse does not
hold in general. In contrast with Nash equilibria, correlated
equilibria can be found for low-degree graphical games (as well as
other classes of concisely-represented multiplayer games) in
polynomial time~\cite{Pap05}. But, for graphical games it is NP-hard
to find a correlated equilibrium that maximizes total
payoff~\cite{PR}. However, the NP-hardness results apply to more
general games than the one we consider here, in particular the graphs
are not trees. From~\cite{aumann} it is also known that
there exist 2-player, 2-action games for which the expected total
payoff of the best correlated equilibrium is higher than the best
Nash equilibrium, and we discuss this issue further in
Section~\ref{sec:conclusions}.

\section{Preliminaries and Notation}\label{defs}

We consider graphical games in which the underlying graph $G$ is an
$n$-vertex tree, in which each vertex has at most $\Delta$ children.
Each vertex has two actions, which are denoted by~$0$ and~$1$. A mixed
strategy of a player $V$ is represented as a single number $v\in[0,
1]$, which denotes the probability that $V$ selects action~$1$.

For the purposes of the algorithm, the tree is rooted arbitrarily. For
convenience, we assume without loss of generality that the root has a
single child, and that its payoff is independent of the action chosen
by the child.  This can be achieved by first choosing an arbitrary
root of the tree, and then adding a dummy ``parent'' of this root,
giving the new parent a constant payoff function, e.g., $0$.

Given an edge $(V,W)$ of the tree~$G$, and a mixed strategy $w$
for~$W$, let $G_{(V,W),W=w}$ be the instance obtained from~$G$ by (1)
deleting all nodes $Z$ which are separated from~$V$ by $W$ (i.e., all
nodes $Z$ such that the path from $Z$ to $V$ passes through $W$), and
(2) restricting the instance so that $W$ is required to play mixed
strategy~$w$.

\begin{definition}
\label{def:bestresponse} Suppose that $(V,W)$ is an edge of the tree,
that $v$ is a mixed strategy for~$V$ and that $w$ is a mixed strategy
for~$W$. We say that $v$ is a \emph{potential best response} to $w$
(denoted by $v\in\pbr_{V}(w)$) if there is an equilibrium in the
instance $G_{(V,W),W=w}$ in which $V$ has mixed strategy~$v$. We
define the {\em best response policy} for $V$, given $W$, as $\PBR(W,
V)=\{(w, v)\mid v\in\pbr_V(w), w\in[0, 1]\}$.
\end{definition}

The upstream pass of the generic algorithm of~\cite{kls} considers
every node~$V$ (other than the root) and computes the best response
policy for~$V$ given its parent. With the above assumptions about the
root, the downstream pass is straightforward. The root selects a mixed
strategy $w$ for the root~$W$ and a mixed strategy $v\in \PBR(W,V)$
for each child $V$ of~$W$. It instructs each child~$V$ to play~$v$.
The remainder of the downward pass is recursive. When a node $V$ is
instructed by its parent to adopt mixed strategy~$v$, it does the
following for each child $U$ --- It finds a pair $(v,u) \in \PBR(V,U)$
(with the same $v$ value that it was given by its parent) and
instructs $U$ to play $u$.

The best response policy for a vertex~$U$ given its parent~$V$ can be
represented as a union of rectangles, where a {\em rectangle} is
defined by a pair of closed intervals $(I_V, I_U)$ and consists of all
points in $I_V\times I_U$; it may be the case that one or both of the
intervals $I_V$ and $I_U$ consists of a single point.  In order to
perform computations on $\PBR(V,U)$, and to bound the number of
rectangles, \cite{egg} used the notion of an \emph{event point}, which
is defined as follows. For any set $A\subseteq[0, 1]^2$ that is
represented as a union of a finite number of rectangles, we say that a
point $u\in[0, 1]$ on the $U$-axis is a $U$-{\em event point} of $A$
if $u=0$ or $u=1$ or the representation of $A$ contains a rectangle of
the form $I_V\times I_U$ and $u$ is an endpoint of $I_U$; $V$-event
points are defined similarly.

For many games considered in this paper, the underlying graph is an
$n$-vertex path, i.e., a graph $G=(V, E)$ with the vertex set 
$V=\{V_1, \dots,
V_n\}$ and the edge set 
$E=\{(V_1, V_2), \dots, (V_{n-1}, V_n)\}$.  In~\cite{egg},
it was shown that for such games, the best response policy has only
polynomially-many rectangles.  The proof that the number of rectangles
in $\PBR(V_{j+1}, V_{j})$ is polynomial proceeds by first showing that the
number of event points in $\PBR(V_{j+1}, V_{j})$ cannot exceed the
number of event points in $\PBR(V_{j}, V_{j-1})$ by more than $2$, and
using this fact to bound the number of rectangles in $\PBR(V_{j+1},
V_{j})$.

Let $P^0(V)$ and $P^1(V)$ be the expected payoffs to $V$ when it plays
0 and 1, respectively. Both $P^0(V)$ and $P^1(V)$ are multilinear
functions of the strategies of $V$'s neighbors.  In what follows, we
will frequently use the following simple observation.
\begin{claim}\label{clm:any}
For a vertex $V$ with a single child $U$ and parent $W$, given any
$A, B, C, D\in \rationals$, $A', B', C', D'\in \rationals$, one can
select the payoffs to $V$ so that $P^0(V)=Auw+Bu+Cw+D$,
$P^1(V)=A'uw+B'u+C'w+D'$. Moreover, if all $A$, $B$, $C$, $D$, $A'$,
$B'$, $C'$, $D'$ are integer, the payoffs to $V$ are integer as well.
\end{claim}
\begin{proof}
We will give the proof for $P^0(V)$; the proof for $P^1(V)$ is
similar.  For $i, j=0, 1$, let $P_{ij}$ be the payoff to $V$ when $U$
plays $i$, $V$ plays $0$ and $W$ plays $j$. We have
$P^0(V)=P_{00}(1-u)(1-w)+P_{10}u(1-w)+P_{01}(1-u)w+P_{11}uw$.  We have
to select the values of $P_{ij}$ so that
$P_{00}-P_{10}-P_{01}+P_{11}=A$, $-P_{00}+P_{10}=B$,
$-P_{00}+P_{01}=C$, $P_{00}=D$.  It is easy to see that the unique
solution is given by $P_{00}=D$, $P_{01}=C+D$, $P_{10}=B+D$,
$P_{11}=A+B+C+D$.
\end{proof}

The input to all algorithms considered in this paper includes
the payoff matrices for each player. We assume that all elements
of these matrices are integer. Let $P_{\max}$ be the greatest
absolute value of any element of any payoff matrix. Then the input
consists of at most $n2^{\Delta+1}$ numbers, each of which can be
represented using $\lceil \log P_{\max}\rceil$ bits.

\section{Nash Equilibria That Maximize The Social Welfare: Solutions
in $\reals\setminus\rationals$}\label{sec:rational}

From the point of view of social welfare, the best Nash equilibrium is
the one that maximizes the sum of the players' expected payoffs.
Unfortunately, it turns out that computing such a strategy profile
exactly is not possible: in this section, we show that even if all
players' payoffs are integers, the strategy profile that maximizes the
total payoff may have irrational coordinates; moreover, it may involve
algebraic numbers of an arbitrary degree.

\subsection{Warm-up: quadratic irrationalities}\label{quad}

We start by providing an example of a graphical game on a path of
length 3 with integer payoffs such that in the Nash equilibrium that
maximizes the total payoff, one of the players has a strategy in
$\reals\setminus\rationals$. In the next subsection, we will extend
this example to algebraic numbers of arbitrary degree $n$; to do so,
we have to consider paths of length $O(n)$.

\begin{theorem}
There exists an integer-payoff graphical game $G$ on a 3-vertex path
$UVW$ such that, in any Nash equilibrium of $G$ that maximizes social
welfare, the strategy, $u$, of the player $U$ and the total payoff,
$p$, satisfy $u, p\in\reals \setminus\rationals$. \label{thm:quad}
\end{theorem}
\begin{proof}
The payoffs to the players in $G$ are specified as follows.  The
payoff to $U$ is identically $0$, i.e., $P^0(U)=P^1(U)=0$.  Using
Claim~\ref{clm:any}, we select the payoffs to $V$ so that
$P^0(V)=-uw+3w$ and $P^1(V)=P^0(V)+w(u+2)-(u+1)$, where $u$ and $w$
are the (mixed) strategies of $U$ and $W$, respectively.  It follows
that $V$ is indifferent between playing 0 and 1 if and only if
$w=f(u)=\frac{u+1}{u+2}$.  Observe that for any $u\in[0, 1]$ we have
$f(u)\in[0, 1]$.  The payoff to $W$ is 0 if it selects the same action
as $V$ and 1 otherwise.

\begin{claim}\label{triple}
All Nash equilibria of the game $G$ are of the form $(u, 1/2, f(u))$.
That is, in any Nash equilibrium, $V$ plays $v=1/2$ and $W$ plays
$w=f(u)$.  Moreover, for any value of $u$, the vector of strategies
$(u, 1/2, f(u))$ constitutes a Nash equilibrium.
\end{claim}
\begin{proof}
It is easy to check that for any $u\in[0, 1]$, the vector $(u, 1/2,
f(u))$ is a Nash equilibrium.  Indeed, $U$ is content to play any
mixed strategy $u$ no matter what $V$ and $W$ do. Furthermore, $V$ is
indifferent between 0 and 1 as long as $w=f(u)$, so it can play $1/2$.
Finally, if $V$ plays 0 and 1 with equal probability, $W$ is
indifferent between 0 and 1, so it can play $f(u)$.

Conversely, suppose that $v>1/2$. Then $W$ strictly prefers to play 0,
i.e., $w=0$. Then for $V$ we have $P^1(V)=P^0(V)-(u+1)$, i.e., $P^1(V)
< P^0(V)$, which implies $v=0$, a contradiction.  Similarly, if
$v<1/2$, player $W$ prefers to play 1, so we have $w=1$.  Hence,
$P^1(V)=P^0(V)+(u+2)-(u+1)$, i.e., $P^1(V) > P^0(V)$, which implies
$v=1$, a contradiction. Finally, if $v=1/2$, but $w\neq f(u)$, player
$V$ is not indifferent between 0 and 1, so he would deviate from
playing $1/2$.  This completes the proof of Claim~\ref{triple}.
%These nested proofs are ugly, because the square doesn't come at the end
%of the outer proof. I've added explanation
\end{proof}

By Claim~\ref{triple}, the total payoff in any Nash equilibrium of
this game is a function of $u$. More specifically, the payoff to $U$
is 0, the payoff to $V$ is $-uf(u)+3f(u)$, and the payoff to $W$ is
$1/2$.  Therefore, the Nash equilibrium with the maximum total payoff
corresponds to the value of $u$ that maximizes
$$
g(u)=-u\frac{(u+1)}{u+2}+3\frac{u+1}{u+2}=-\frac{(u-3)(u+1)}{u+2}.
$$
To find extrema of $g(u)$, we compute $h(u)=-\frac{d}{du}g(u)$. We have
$$
h(u)=\frac{(2u-2)(u+2)-(u-3)(u+1)}{(u+2)^2}
=\frac{u^2+4u-1}{(u+2)^2}.
$$
Hence, $h(u)=0$ if and only if $u\in\{-2+\sqrt{5}, -2-\sqrt{5}\}$.
Note that $-2+\sqrt{5}\in[0, 1]$.

The function $g(u)$ changes sign at $-2$, $-1$, and $3$. We have
$g(u)<0$ for $g>3$, $g(u)>0$ for $u<-2$, so the extremum of $g(u)$
that lies between $1$ and $3$, i.e., $u=-2+\sqrt{5}$, is a local
maximum. We conclude that the social welfare-maximizing Nash
equilibrium for this game is given by the vector of strategies
$(-2+\sqrt{5}, 1/2, (5-\sqrt{5})/5)$. The respective total payoff is
$$
0-\frac{(\sqrt{5}-5)(\sqrt{5}-1)}{\sqrt{5}}+\frac{1}{2}=13/2-2\sqrt{5}.
$$
This concludes the proof of Theorem~\ref{thm:quad}.
\end{proof}

\subsection{Strategies of arbitrary degree}\label{algebraic}

We have shown that in the social welfare-maximizing Nash equilibrium,
some players' strategies can be quadratic irrationalities, and so can
the total payoff.  In this subsection, we will extend this result to
show that we can construct an integer-payoff graphical game on a path
whose social welfare-maximizing Nash equilibrium involves arbitrary
algebraic numbers in $[0, 1]$.

\begin{theorem}
For any degree-$n$ algebraic number $\alpha\in[0, 1]$, there exists an
integer payoff graphical game on a path of length $O(n)$ such that, in
all social welfare-maximizing Nash equilibria of this game, one of the
players plays $\alpha$.
\label{thm:arbitrary}
\end{theorem}
\begin{proof}
Our proof consists of two steps. First, we construct a rational
expression $R(x)$ and a segment $[x', x'']$ such that $x', x''\in
\rationals$ and $\alpha$ is the only maximum of $R(x)$ on $[x', x'']$.
Second, we construct a graphical game whose Nash equilibria can be
parameterized by $u\in[x', x'']$, so that at the equilibrium that
corresponds to $u$ the total payoff is $R(u)$ and, moreover, 
some player's strategy is $u$. 
It follows that to achieve the payoff-maximizing Nash equilibrium, this
player has to play $\alpha$.  The details follow.

\begin{lemma}\label{alg}
Given an algebraic number $\alpha\in[0, 1]$,
$\deg(\alpha)=n$, there exist
$K_2, \dots, K_{2n+2}\in\rationals$ and
$x', x''\in(0, 1)\cap\rationals$ such that
$\alpha$ is the only maximum of
$$
R(x)=\frac{K_2}{x+2}+\cdots+\frac{K_{2n+2}}{x+2n+2}
$$
on $[x', x'']$.
\end{lemma}
\begin{proof}
Let $P(x)$ be the minimal polynomial of $\alpha$, i.e., a polynomial
of degree $n$ with rational coefficients whose leading coefficient is 1
such that $P(\alpha)=0$.
Let $A=\{\alpha_1, \dots, \alpha_n\}$ be the set of all roots of
$P(x)$.
Consider the polynomial $Q_1(x)=-P^2(x)$.
It has the same roots as $P(x)$, and moreover, for any
$x\not\in A$ we have $Q_1(x)<0$.
Hence, $A$ is the set of all maxima of $Q_1(x)$.
Now, set
$R(x)=\frac{Q_1(x)}{(x+2)\dots(x+2n+1)(x+2n+2)}$.
Observe that $R(x)\le 0$ for all $x\in[0, 1]$ and $R(x)=0$
if and only if $Q_1(x)=0$. Hence, the set $A$
is also the set of all maxima of $R(x)$ on $[0, 1]$.

Let $d=\min\{|\alpha_i-\alpha|\mid
                 \alpha_i\in A, \alpha_i\neq\alpha\}$, and set
$\alpha'=\max\{\alpha-d/2, 0\}$,
$\alpha''=\min\{\alpha+d/2, 1\}$.
Clearly, $\alpha$ is the only zero
(and hence, the only maximum) of $R(x)$ on $[\alpha', \alpha'']$.
Let $x'$ and $x''$ be some rational numbers in $(\alpha', \alpha)$ and
$(\alpha, \alpha'')$, respectively; note that by excluding
the endpoints of the intervals we ensure that $x', x''\neq 0, 1$.
As $[x', x'']\subset[\alpha', \alpha'']$,
we have that $\alpha$ is the only maximum
of $R(x)$ on $[x', x'']$.

As $R(x)$ is a proper rational expression and all roots
of its denominator are simple,
by partial fraction decomposition theorem, $R(x)$ can be represented
as
$$
R(x)=\frac{K_2}{x+2}+\cdots+\frac{K_{2n+2}}{x+2n+2},
$$
where $K_2, \dots, K_{2n+2}$ are rational numbers.
\end{proof}

Consider a graphical game on the path
$$
U_{-1}V_{-1}U_0V_0U_1V_1\dots U_{k-1}V_{k-1}U_k,
$$
where $k=2n+2$.  Intuitively, we want each triple $(U_{i-1},
V_{i-1}, \linebreak[3]U_i)$ to behave similarly to the players $U$,
$V$, and $W$ from the game described in the previous subsection. More
precisely, we define the payoffs to the players in the following way.
\begin{itemize}
\item
The payoff to $U_{-1}$ is $0$ no matter what everyone else does.
\item
The expected payoff to $V_{-1}$ is $0$ if it plays $0$ and
$u_0-(x''-x')u_{-1}-x'$ if it plays $1$,
where $u_0$ and $u_{-1}$ are the strategies of
$U_0$ and $U_{-1}$, respectively.
\item
The expected payoff to $V_0$ is $0$ if it plays $0$ and
$u_1(u_0+1)-u_0$ if it plays $1$,
where $u_0$ and $u_1$ are the strategies of
$U_0$ and $U_1$, respectively.
\item
For each $i=1, \dots, k-1$, the expected payoff to $V_i$ when it plays
0 is $P^0(V_i)=A_iu_iu_{i+1}-A_iu_{i+1}$,
and the expected payoff to $V_i$ when it plays
1 is $P^1(V_i)=P^0(V_i)+u_{i+1}(2-u_i)-1$,
where $A_i=-K_{i+1}$  and $u_{i+1}$ and $u_i$ are
the strategies of $U_{i+1}$ and $U_i$, respectively.
\item
For each $i=0, \dots, k$, the payoff to $U_i$ does not depend on
$V_i$ and is $1$ if $U_i$ and $V_{i-1}$ select different actions and
$0$ otherwise.
\end{itemize}
We will now characterize the Nash equilibria of this game using
a sequence of claims.

\begin{claim}
In all Nash equilibria of this game $V_{-1}$ plays $1/2$, and the
strategies of $u_{-1}$ and $u_0$ satisfy
$u_0=(x''-x')u_{-1}+x'$. Consequently, in all Nash equilibria we have
$u_0\in[x', x'']$.
\end{claim}

\begin{proof}
The proof is similar to that of Claim~\ref{triple}.  Let
$f(u_{-1})=(x''-x')u_{-1}+x'$. Clearly, the player $V_{-1}$ is
indifferent between playing 0 and 1 if and only if $u_0=f(u_{-1})$.
Suppose that $v_{-1}<1/2$. Then $U_0$ strictly prefers to play 1,
i.e., $u_0=1$, so we have
$$
P^1(V_{-1})=P^0(V_{-1})+1-(x''-x')u_{-1}-x'.
$$
As
$$
1-x''\le 1-(x''-x')u_{-1}-x'\le 1-x'
$$
for $u_{-1}\in[0, 1]$ and $x''<1$,
we have
$P^1(V_{-1})>P^0(V_{-1})$, so $V_{-1}$ prefers to play 1, a contradiction.
Similarly, if $v_{-1}>1/2$, the player $U_0$ strictly prefers to play 0,
i.e., $u_0=0$, so we have
$$
P^1(V_{-1})=P^0(V_{-1})-(x''-x')u_{-1}-x'.
$$ As $x'<x''$, $x'>0$, we have $P^1(V_{-1})<P^0(V_{-1})$, so $V_{-1}$
prefers to play 0, a contradiction.  Finally, if $V_{-1}$ plays $1/2$,
but $u_0\neq f(u_{-1})$, player $V_{-1}$ is not indifferent between 0
and 1, so he would deviate from playing $1/2$.

Also, note that $f(0)=x'$, $f(1)=x''$, and, moreover,
$f(u_{-1})\in[x', x'']$ if and only if $u_{-1}\in[0, 1]$. Hence, in
all Nash equilibria of this game we have $u_0\in[x', x'']$.
\end{proof}

\begin{claim}
In all Nash equilibria of this game for each $i=0, \dots, k-1$, we
have $v_i=1/2$, and the strategies of the players $U_i$ and $U_{i+1}$
satisfy $u_{i+1}=f_i(u_i)$, where $f_0(u)=u/(u+1)$ and
$f_i(u)=1/(2-u)$ for $i>0$.
\end{claim}

\begin{proof}
The proof of this claim is also similar to that of Claim~\ref{triple}.
We use induction on $i$ to prove that the statement of the claim is
true and, additionally, $u_i\neq 1$ for $i>0$.

For the base case $i=0$, note that $u_0\neq 0$ by the previous claim
(recall that $x'$, $x''$ are selected so that $x', x''\neq 0, 1$) and
consider the triple $(U_0, V_0, U_1)$.  Let $v_0$ be the strategy of
$V_0$.  First, suppose that $v_0>1/2$. Then $U_1$ strictly prefers to
play 0, i.e., $u_1=0$.  Then for $V_0$ we have
$P^1(V_0)=P^0(V_0)-u_0$. As $u_0\neq 0$, we have $P^1(V_0) <
P^0(V_0)$, which implies $v_1=0$, a contradiction.  Similarly, if
$v_0<1/2$, player $U_1$ prefers to play 1, so we have $u_1=1$.  Hence,
$P^1(V_0)=P^0(V_0)+1$. It follows that $P^1(V_0) > P^0(V_0)$, which
implies $v_0=1$, a contradiction.  Finally, if $v_0=1/2$, but $u_1\neq
u_0/(u_0+1)$, player $V_0$ is not indifferent between 0 and 1, so he
would deviate from playing $1/2$. Moreover, as $u_1=u_0/(u_0+1)$ and
$u_0\in[0, 1]$, we have $u_1\neq 1$.

The argument for the inductive step is similar.
Namely, suppose that the statement is proved for all $i'<i$
and consider the triple $(U_i, V_i, U_{i+1})$.

Let $v_i$ be the strategy of $V_i$.  First, suppose that
$v_i>1/2$. Then $U_{i+1}$ strictly prefers to play 0, i.e.,
$u_{i+1}=0$.  Then for $V_i$ we have $P^1(V_i)=P^0(V_i)-1$, i.e.,
$P^1(V_i) < P^0(V_i)$, which implies $v_i=0$, a contradiction.
Similarly, if $v_i<1/2$, player $U_{i+1}$ prefers to play 1, so we
have $u_{i+1}=1$.  Hence, $P^1(V_i)=P^0(V_i)+1-u_i$. By inductive
hypothesis, we have $u_i<1$. Consequently, $P^1(V_i) > P^0(V_i)$,
which implies $v_i=1$, a contradiction.  Finally, if $v_i=1/2$, but
$u_{i+1}\neq 1/(2-u_i)$, player $V_i$ is not indifferent between 0 and
1, so he would deviate from playing $1/2$. Moreover, as
$u_{i+1}=1/(2-u_i)$ and $u_i<1$, we have $u_{i+1}<1$.
\end{proof}

\begin{claim}\label{searchspace}
Any strategy profile of the form
$$
(u_{-1}, 1/2, u_0, 1/2, u_1, 1/2, \dots, u_{k-1}, 1/2, u_k),
$$
where
$u_{-1}\in[0, 1]$,
$u_0=(x''-x')u_{-1}+x'$, $u_1=u_0/(u_0+1)$, and $u_{i+1}=1/(2-u_i)$
for $i\ge 1$ constitutes a Nash equilibrium.
\end{claim}
\begin{proof}
First, the player $U_{-1}$'s payoffs do not depend on other players'
actions, so he is free to play any strategy in $[0, 1]$.
As long as $u_0=(x''-x')u_{-1}+x'$, player $V_{-1}$ is indifferent between
0 and 1, so he is content to play $1/2$; a similar argument applies
to players $V_0, \dots, V_{k-1}$. Finally, for each $i=0, \dots, k$,
the payoffs of player $U_i$ only depend on the strategy of player
$V_{i-1}$. In particular, as long as $v_{i-1}=1/2$, player $U_i$
is indifferent between playing 0 and 1, so he can play any mixed
strategy $u_i\in[0, 1]$. To complete the proof, note that
$(x''-x')u_{-1}+x'\in[0, 1]$ for all $u_{-1}\in[0, 1]$,
$u_0/(u_0+1)\in[0, 1]$ for all $u_0\in[0, 1]$, and
$1/(2-u_i)\in[0, 1]$ for all $u_i\in[0, 1]$, so we have
$u_i\in[0, 1]$ for all $i=0, \dots, k$.
\end{proof}

Now, let us compute the total payoff under a strategy profile of the
form given in Claim~\ref{searchspace}.  The payoff to $U_{-1}$ is 0,
and the expected payoff to each of the $U_i$, $i=0, \dots, k$, is
$1/2$.  The expected payoffs to $V_{-1}$ and $V_0$ are 0.  Finally,
for any $i=1, \dots, k-1$, the expected payoff to $V_i$ is
$T_i=A_iu_iu_{i+1}-A_iu_{i+1}$.  It follows that to find a Nash
equilibrium with the highest total payoff, we have to maximize
$\sum_{i=1}^{k-1}T_i$ subject to conditions $u_{-1}\in[0, 1]$,
$u_0=(x''-x')u_{-1}+x'$, $u_1=u_0/(u_0+1)$, and $u_{i+1}=1/(2-u_i)$
for $i=1, \dots, k-1$.

We would like to express $\sum_{i=1}^{k-1}T_i$ as a function of
$u_0$. To simplify notation, set $u=u_0$.
\begin{lemma}
For $i=1, \dots, k$, we have $u_i=\frac{u+i-1}{u+i}$.
\end{lemma}
\begin{proof}
The proof is by induction on $i$.
For $i=1$, we have $u_1=u/(u+1)$.
Now, for $i\ge 2$ suppose that $u_{i-1}=(u+i-2)/(u+i-1)$.
We have
$u_i=1/(2-u_{i-1})=(u+i-1)/(2u+2i-2-u-i+2)=(u+i-1)/(u+i)$.
\end{proof}
It follows that for $i=1, \dots, k-1$ we have
\begin{eqnarray*}
T_i=A_i\frac{u+i-1}{u+i}\frac{u+i}{u+i+1}-A_i\frac{u+i}{u+i+1} =\\
-A_i\frac{1}{u+i+1}=\frac{K_{i+1}}{u+i+1}.
\end{eqnarray*}

Observe that as $u_{-1}$ varies from 0 to 1,
$u$ varies from $x'$ to $x''$. Therefore, to maximize the total
payoff, we have to choose $u\in[x', x'']$ so as to maximize
$$
\frac{K_2}{u+2}+\dots+\frac{K_k}{u+k}=R(u).
$$
By construction, the only maximum of $R(u)$ on $[x', x'']$
is $\alpha$. It follows that in the payoff-maximizing
Nash equilibrium of our game $U_0$ plays $\alpha$.

Finally, note that the payoffs in our game are rational
rather than integer. However, it is easy to see that we can multiply
all payoffs to a player by their greatest common denominator without
affecting his strategy. In the resulting game, all payoffs are integer.
This concludes the proof of Theorem~\ref{thm:arbitrary}.

\end{proof}
%Moreover, one can show that the payoffs do not have to be very large,
%namely, each payoff can be represented with $\poly(2^{\poly n}, \log |C|)$
%bits, where $C$ is the coefficient of the minimal polynomial
%$P(x)$ that has the largest absolute value.

\section{Approximating The Socially Optimal Nash Equilibrium}\label{bestnash}

We have seen that the Nash equilibrium that maximizes the social
welfare may involve strategies that are not in $\rationals$. Hence, in
this section we focus on finding a Nash equilibrium that is {\em
almost} optimal from the social welfare perspective.  We propose an
algorithm that for any $\epsilon>0$ finds a Nash equilibrium whose
total payoff is within $\epsilon$ from optimal. The running time of
this algorithm is polynomial in $1/\epsilon$, $n$ and $|P_{\max}|$
(recall that $P_{\max}$ is the maximum absolute value of an entry
of a payoff matrix).

While the negative result of the previous section is for graphical
games on paths, our algorithm applies to a wider range of scenarios.
Namely, it runs in polynomial time on bounded-degree trees as long as
the best response policy of each vertex, given its parent, can be
represented as a union of a polynomial number of rectangles.  Note
that path graphs always satisfy this condition: in~\cite{egg} we
showed how to compute such a representation, given a graph with
maximum degree~2. Consequently, for path graphs the running time of
our algorithm is guaranteed to be polynomial.
(Note that~\cite{egg} exhibits a family of graphical games on
bounded-degree trees for which the best response policies of some of
the vertices, given their parents, have exponential size, when
represented as unions of rectangles.)

Due to space restrictions, in this version of the paper 
we present the algorithm for the case where the graph 
underlying the graphical game is a path.
We then state our result for the general case; the proof
can be found in the appendix.

Suppose that $\s$ is a strategy profile for a graphical game~$G$.
That is, $\s$ assigns a mixed strategy to each vertex of~$G$.
let $EP_V(\s)$ be the expected payoff of player~$V$ under $\s$
and let $EP(\s) = \sum_V EP_V(\s)$.
Let $$M(G) = \max\{EP(\s) \mid \s \mbox{ is a Nash equilibrium for $G$}\}.$$

\begin{theorem}\label{thm:bestnash}
Suppose that $G$ is a graphical game on an $n$-vertex path.  Then for
any $\epsilon>0$ there is an algorithm that constructs a Nash
equilibrium $\s'$ for $G$ that satisfies $EP(\s') \geq
M(G)-\epsilon$. The running time of the algorithm is
$O(n^4P_{\max}^3/\epsilon^3)$
\end{theorem}

\begin{proof}
Let $\{V_1, \dots, V_n\}$ be the set of all players.  We start by
constructing the best response policies for all $V_i$, $i=1, \dots,
n-1$. As shown in~\cite{egg}, this can be done in time $O(n^3)$.

Let $N>5n$ be a parameter to be selected later, set $\delta=1/N$, and
define
%don't use G for that, G is the graphical game!
$X=\{j\delta \mid j=0, \dots, N\}$.  We say that $v_j$ is an event
point for a player $V_i$ if it is a $V_i$-event point for $\PBR(V_{i},
V_{i-1})$ or $\PBR(V_{i+1}, V_{i})$.  For each player $V_i$, consider
a finite set of strategies $X_i$ given by
\begin{align*}
X_i=X\cup
\{v_j|v_j\text{ is an event point for }V_{i}\}.
\end{align*}
It has been shown in~\cite{egg} that for any
$i=2, \dots, n$, the best response policy $\PBR(V_{i}, V_{i-1})$
has at most $2n+4$ $V_i$-event points.
As we require $N>5n$, we have $|X_i|\le 2N$;
assume without loss of generality that $|X_i|=2N$.
Order the elements of $X_i$ in increasing order as
$x_i^1=0<x_i^2<\dots<x_i^{2N}$.
We will refer to the strategies in $X_i$ as {\em discrete} strategies
of player $V_i$; a strategy profile in which each player has a discrete
strategy will be referred to as a {\em discrete} strategy profile.

We will now show that even we restrict each player $V_i$ to strategies
from $X_i$, the players can still achieve a Nash equilibrium, and
moreover, the best such Nash equilibrium (with respect to the social
welfare) has total payoff at least $M(G)-\epsilon$ as long as $N$ is
large enough.

Let $\s$ be a strategy profile that maximizes social welfare.  That
is, let $\s=(s_1, \dots, s_n)$ where $s_i$ is the mixed strategy of
player $V_i$ and $EP(\s)=M(G)$.  For $i=1, \dots, n$, let
$t_i=\max\{x_{i}^j\mid x_i^j\le s_i\}$.  First, we will show that the
strategy profile $\t=(t_1, \dots, t_n)$ is a Nash equilibrium for $G$.

Fix any $i$, $1<i\le n$, and let $R=[v_1, v_2]{\times}[u_1, u_2]$ be
the rectangle in $\PBR(V_{i}, V_{i-1})$ that contains $(s_i,
s_{i-1})$.  As $v_1$ is a $V_i$-event point of $\PBR(V_{i}, V_{i-1})$,
we have $v_1\le t_i$, so the point $(t_i, s_{i-1})$ is inside
$R$. Similarly, the point $u_1$ is a $V_{i-1}$-event point of
$\PBR(V_{i}, V_{i-1})$, so we have $u_1\le t_{i-1}$, and therefore the
point $(t_{i}, t_{i-1})$ is inside $R$.  This means that for any $i$,
$1<i\le n$, we have $t_{i-1}\in\pbr_{V_{i-1}}(t_i)$, which implies
that $\t=(t_1, \dots, t_n)$ is a Nash equilibrium for $G$.

Now, let us estimate the expected loss in social welfare
caused by playing $\t$ instead of $\s$.
\begin{lemma}\label{3delta}
For any pair of strategy profiles $\t, \s$ such that
$|t_i-s_i|\le\delta$ we have $|EP_{V_i}(\s)-EP_{V_i}(\t)|\le
24P_{\max}\delta$ for any $i=1, \dots, n$.
\end{lemma}
\begin{proof}
Let $P_{klm}^i$ be the payoff of the player $V_i$, when he plays
$k$, $V_{i-1}$ plays $l$, and $V_{i+1}$ plays $m$.
Fix $i=1, \dots, n$ and for $k, l, m\in\{0, 1\}$, set
\begin{eqnarray*}
t^{klm} & = &
t_{i-1}^k(1-t_{i-1})^{1-k}t_{i}^l(1-t_{i})^{1-l}t_{i+1}^m(1-t_{i+1})^{1-m}\\
s^{klm} & = &
s_{i-1}^k(1-s_{i-1})^{1-k}s_{i}^l(1-s_{i})^{1-l}s_{i+1}^m(1-s_{i+1})^{1-m}.
\end{eqnarray*}
We have
\begin{eqnarray*}
|EP_{V_i}(\s)-EP_{V_i}(\t)|\le \sum_{k, l, m=0, 1}|P^i_{klm}(t^{klm}-s^{klm})|\le\\
8P_{\max}\max_{klm}|t^{klm}-s^{klm}|
\end{eqnarray*}
We will now show that
for any $k, l, m\in\{0, 1\}$ we have $|t^{klm}-s^{klm}|\le 3\delta$;
clearly, this implies the lemma.

Indeed, fix $k, l, m\in\{0, 1\}$.
Set
\begin{align*}
x & = &t_{i-1}^k(1-t_{i-1})^{1-k},
x' & = &s_{i-1}^k(1-s_{i-1})^{1-k},\\
y & = &t_{i}^l(1-t_{i})^{1-l},
y' & = &s_{i}^l(1-s_{i})^{1-l},\\
z  & = &t_{i+1}^m(1-t_{i+1})^{1-m},
z' &  = &s_{i+1}^m(1-s_{i+1})^{1-m}.
\end{align*}
Observe that if $k=0$ then $x-x'=(1-t_{i-1})-(1-s_{i-1})$, and
if $k=1$ then $x-x'=t_{i-1}-s_{i-1}$, so $|x-x'|\le \delta$.
A similar argument shows $|y-y'|\le\delta$, $|z-z'|\le\delta$.
Also, we have $x, x', y, y', z, z'\in[0, 1]$.
Hence,
$|t^{klm}-s^{klm}|=|xyz-x'y'z'|=|xyz-x'yz+x'yz-x'y'z+x'y'z-x'y'z'|\le
|x-x'|yz+|y-y'|x'z+|z-z'|x'y'\le 3\delta$.
\end{proof}

Lemma~\ref{3delta} implies $\sum_{i=1}^n|EP_{V_i}(\s)-EP_{V_i}(\t)|\le
24nP_{\max}\delta$, so by choosing $\delta<\epsilon/(24nP_{\max})$,
or, equivalently, setting $N>24nP_{\max}/\epsilon$, we can ensure that
the total expected payoff for the strategy profile $\t$ is within
$\epsilon$ from optimal.

We will now show that we can find the best discrete Nash equilibrium
(with respect to the social welfare) using dynamic programming.  As
$\t$ is a discrete strategy profile, this means that the strategy
profile found by our algorithm will be at least as good as $\t$.

Define $m_i^{l, k}$ to be the maximum total payoff that $V_1,
\dots, V_{i-1}$ can achieve if each $V_j$, $j\le i$, chooses a
strategy from $X_j$, for each $j<i$ the strategy of $V_j$ is a
potential best response to the strategy of $V_{j+1}$, and, moreover,
$V_{i-1}$ plays $x_{i-1}^l$, $V_i$ plays $x_i^k$.  If there is no way
to choose the strategies for $V_1, \dots, V_{i-1}$ to satisfy these
conditions, we set $m_i^{l, k}=-\infty$.  The values $m_i^{l, k}$,
$i=1, \dots, n$; $k, l=1, \dots, N$, can be computed inductively, as
follows.

We have $m_1^{l, k}=0$ for $k, l=1, \dots, N$.  Now, suppose that we
have already computed $m_{j}^{l, k}$ for all $j<i$; $k, l=1, \dots,
N$.  To compute $m_i^{k, l}$, we first check if $(x_i^k,
x_{i-1}^l)\in\PBR (V_{i}, V_{i-1})$. If this is not the case, we have
$m_i^{l, k}=-\infty$.  Otherwise, consider the set
$Y=X_{i-2}\cap\pbr_{V_{i-2}}(x_{i-1}^l)$, i.e., the set of all
discrete strategies of $V_{i-2}$ that are potential best responses to
$x_{i-1}^l$.  The proof of Theorem~1 in~\cite{egg} implies that the
set $\pbr_{V_{i-2}}(x_{i-1}^l)$ is non-empty: the player $V_{i-2}$ has
a potential best response to any strategy of $V_{i-1}$, in particular,
$x_{i-1}^l$.  By construction of the set $X_{i-2}$, this implies that
$Y$ is not empty.  For each $x_{i-2}^j\in Y$, let $p_{jlk}$ be the
payoff that $V_{i-1}$ receives when $V_{i-2}$ plays $x_{i-2}^j$,
$V_{i-1}$ plays $x_{i-1}^l$, and $V_i$ plays $x_i^k$. Clearly,
$p_{jlk}$ can be computed in constant time. Then we have $m_i^{l,
k}=\max\{m_{i-1}^{j, l}+p_{jlk}\mid x_{i-2}^j\in Y\}$.

Finally, suppose that we have computed $m_n^{l, k}$ for $l, k=1,
\dots, N$.  We still need to take into account the payoff of player
$V_n$.  Hence, we consider all pairs $(x_n^k, x_{n-1}^l)$ that
satisfy $x_{n-1}^l\in\pbr_{V_{n-1}}(x_n^k)$, and pick the one that
maximizes the sum of $m_n^{k, l}$ and the payoff of $V_n$ when he
plays $x_n^k$ and $V_{n-1}$ plays $x_{n-1}^l$. This results in the
maximum total payoff the players can achieve in a Nash equilibrium
using discrete strategies; the actual strategy profile that produces
this payoff can be reconstructed using standard dynamic programming
techniques.

It is easy to see that each $m_i^{l, k}$ can be computed in time
$O(N)$, i.e., all of them can be computed in time $O(nN^3)$. Recall
that we have to select $N\ge (24nP_{\max})/\epsilon$ to ensure that
the strategy profile we output has total payoff that is within
$\epsilon$ from optimal. We conclude that we can compute an
$\epsilon$-approximation to the best Nash equilibrium in time
$O(n^4P_{\max}^3/\epsilon^3)$.  This completes the proof of
Theorem~\ref{thm:bestnash}.
\end{proof}

To state our result for the general case (i.e., when the underlying
graph is a bounded-degree tree rather than a path), we need additional
notation.  If $G$ has $n$ players, let $q(n)$ be an upper bound on the
number of event points in the representation of any best response
policy. That is, we assume that for any vertex $U$ with parent~$V$,
$B(V,U)$ has at most $q(n)$ event points. We will be interested in the
situation in which $q(n)$ is polynomial in~$n$.

\begin{theorem}
\label{cor:tree} Let $G$ be an $n$-player graphical game on a tree in which
each node has at most~$\Delta$ children. Suppose we are given a set of
best-response policies for~$G$ in which each best-response
policy~$B(V,U)$ is represented by a set of rectangles with at most
$q(n)$ event points. For any $\epsilon>0$, there is an algorithm that
constructs a Nash equilibrium~$\s'$ for~$G$ that satisfies $EP(\s')
\ge M(G) - \epsilon$. The running time of the algorithm is polynomial
in $n$, $P_{\mathrm{max}}$ and $\epsilon^{-1}$ provided that the tree
has bounded degree (that is, $\Delta=O(1)$) and $q(n)$ is a polynomial
in~$n$. In particular, if
$$N=\max((\Delta+1)q(n)+1,
n 2^{\Delta+2}(\Delta+2)P_{\max}\epsilon^{-1}
)$$ and $\Delta>1$
then the
running time is $O(n \Delta {(2N)}^\Delta$.
\end{theorem}
The proof of this theorem can be found in the appendix.

\subsection{A polynomial-time algorithm for multiplicative approximation}
\label{multiplicative}

The running time of our algorithm is pseudopolynomial rather than
polynomial, because it includes a factor which is polynomial in
$P_{\mathrm{max}}$, the maximum (in absolute value) entry in any
payoff matrix.  If we are interested in multiplicative approximation
rather than additive one, this can be improved to polynomial.

First, note that we cannot expect a multiplicative approximation for
all inputs.  That is, we cannot hope to have an algorithm that
computes a Nash equilibrium with total payoff at least
$(1-\epsilon)M(G)$. If we had such an algorithm, then for graphical
games $G$ with $M(G)=0$, the algorithm would be required to output the
optimal solution.  To show that this is infeasible, observe that we
can use the techniques of Section~\ref{algebraic} to construct two
integer-coefficient graphical games on paths of length $O(n)$ such
that for some $X\in\reals$ the maximal total payoff in the first game
is $X$, the maximal total payoff in the second game is $-X$, and for
both games, the strategy profiles that achieve the maximal total
payoffs involve algebraic numbers of degree $n$. By combining the two
games so that the first vertex of the second game becomes connected to
the last vertex of the first game, but the payoffs of all players do
not change, we obtain a graphical game in which the best Nash
equilibrium has total payoff $0$, yet the strategies that lead to this
payoff have high algebraic complexity.

However, we can achieve a multiplicative approximation when all
entries of the payoff matrices are positive and the ratio between any
two entries is polynomially bounded.  Recall that we assume that all
payoffs are integer, and let $P_{\min}>0$ be the smallest entry of any
payoff matrix.  In this case, for any strategy profile the payoff to
player $i$ is at least $P_{\min}$, so the total payoff in the
social-welfare maximizing Nash equilibrium $\s$ satisfies $M(G)\ge
nP_{\min}$.  Moreover, Lemma~\ref{3delta} implies that by choosing
$\delta<\epsilon/(24P_{\max}/P_{\min})$, we can ensure that the Nash
equilibrium $\t$ produced by our algorithm satisfies
$$
\sum_{i=1}^nEP_{V_i}(\s)-\sum_{i=1}^nEP_{V_i}(\t)\le
24P_{\max}\delta n\le\epsilon nP_{\min}\le \epsilon M(G),
$$ i.e., for this value of $\delta$ we have
$\sum_{i=1}^nEP_{V_i}(\t)\ge (1-\epsilon)M(G)$. Recall that the
running time of our algorithm is $O(nN^3)$, where $N$ has to be
selected to satisfy $N>5n$, $N=1/\delta$.  It follows that if
$P_{\min}>0$, $P_{\max}/P_{\min}=\poly(n)$, we can choose $N$ so that
our algorithm provides a multiplicative approximation guarantee and
runs in time polynomial in $n$ and $1/\epsilon$.

\section{Bounded Payoff Nash Equilibria}\label{bounds}

Another natural way to define what is a ``good'' Nash equilibrium is
to require that each player's expected payoff exceeds a certain
threshold.  These thresholds do not have to be the same for all
players.  In this case, in addition to the payoff matrices of the $n$
players, we are given $n$ numbers $T_1, \dots, T_n$, and our goal is
to find a Nash equilibrium in which the payoff of player $i$ is at
least $T_i$, or report that no such Nash equilibrium exists.  It turns
out that we can design an FPTAS for this problem using the same
techniques as in the previous section.

\begin{theorem}
Given a graphical game $G$ on an $n$-vertex path and $n$ rational
numbers $T_1, \dots, T_n$, suppose that there exists a strategy
profile $\s$ such that $\s$ is a Nash equilibrium for $G$ and
$EP_{V_i}(\s)\ge T_i$ for $i=1, \dots, n$.  Then for any $\epsilon>0$
we can find in time $O(\max\{nP_{\max}^3/\epsilon^3,
n^4/\epsilon^3\})$ a strategy profile $\s'$ such that $\s'$ is a Nash
equilibrium for $G$ and $EP_{V_i}(\s')\ge T_i-\epsilon$ for $i=1,
\dots, n$.
\label{thm:besteach}
\end{theorem}
\begin{proof}
The proof is similar to that of Theorem~\ref{thm:bestnash}.
First, we construct the best response policies for all players,
choose $N>5n$, and construct the sets $X_i$, $i=1, \dots, n$,
as described in the proof of Theorem~\ref{thm:bestnash}.

Consider a strategy profile $\s$ such that $\s$ is a Nash equilibrium
for $G$ and $EP_{V_i}(\s)\ge T_i$ for $i=1, \dots, n$.  We construct a
strategy profile $t_i=\max\{x_{i}^j\mid x_i^j\le s_i\}$ and use the
same argument as in the proof of Theorem~\ref{thm:bestnash} to show
that $\t$ is a Nash equilibrium for $G$.  By Lemma~\ref{3delta}, we
have $|EP_{V_i}(\s)-EP_{V_i}(\t)|\le 24P_{\max}\delta$, so choosing
$\delta<\epsilon/(24P_{\max})$, or, equivalently, $N>\max\{5n,
24P_{\max}/\epsilon\}$, we can ensure $EP_{V_i}(\t)\ge T_i-\epsilon$
for $i=1, \dots, n$.

Now, we will use dynamic programming to find a discrete Nash
equilibrium that satisfies $EP_{V_i}(\t)\ge T_i-\epsilon$ for $i=1,
\dots, n$. As $\t$ is a discrete strategy profile, our algorithm will
succeed whenever there is a strategy profile $\s$ with
$EP_{V_i}(\s)\ge T_i-\epsilon$ for $i=1, \dots, n$.

Let $z_i^{l, k}=1$ if there is a discrete strategy profile such that
for any $j<i$ the strategy of the player $V_j$ is a potential best
response to the strategy of $V_{j+1}$, the expected payoff of $V_j$ is
at least $T_j-\epsilon$, and, moreover, $V_{i-1}$ plays $x_{i-1}^l$,
$V_i$ plays $x_i^k$.  Otherwise, let $z_i^{l, k}=0$.  We can compute
$z_i^{l, k}$, $i=1, \dots, n$; $k, l=1, \dots, N$ inductively, as
follows.

We have $z_1^{l, k}=1$ for $k, l=1, \dots, N$.  Now, suppose that we
have already computed $z_{j}^{l, k}$ for all $j<i$; $k, l=1, \dots,
N$.  To compute $z_i^{k, l}$, we first check if $(x_i^k,
x_{i-1}^l)\in\PBR (V_{i}, V_{i-1})$.  If this is not the case,
clearly, $z_i^{k, l}=0$.  Otherwise, consider the set
$Y=X_{i-2}\cap\pbr_{V_{i-2}}(x_{i-1}^l)$, i.e., the set of all
discrete strategies of $V_{i-2}$ that are potential best responses to
$x_{i-1}^l$. It has been shown in the proof of
Theorem~\ref{thm:bestnash} that $Y\neq\emptyset$.  For each
$x_{i-2}^j\in Y$, let $p_{jlk}$ be the payoff that $V_{i-1}$ receives
when $V_{i-2}$ plays $x_{i-2}^j$, $V_{i-1}$ plays $x_{i-1}^l$, and
$V_i$ plays $x_i^k$. Clearly, $p_{jlk}$ can be computed in constant
time.  If there exists an $x_{i-2}^j\in Y$ such that $z_{i-1}^{j,
l}=1$ and $p_{jlk}\ge T_{i-2}-\epsilon$, set $z_{i}^{l, k}=1$.
Otherwise, set $z_i^{l, k}=0$.

Having computed $z_n^{l, k}$, $l, k=1, \dots, N$, we check if $z_n^{l,
k}=1$ for some pair $(l, k)$. if such a pair of indices exists, we
instruct $V_n$ to play $x_n^k$ and use dynamic programming techniques
(or, equivalently, the downstream pass of the algorithm of~\cite{kls})
to find a Nash equilibrium $\s'$ that satisfies $EP_{V_i}(\s')\ge
T_i-\epsilon$ for $i=1, \dots, n$ (recall that $V_n$ is a dummy
player, i.e., we assume $T_n=0$, $EP_n(\s')=0$ for any choice of
$\s'$).  If $z_n^{l, k}=0$ for all $l, k=1, \dots, N$, there is no
discrete Nash equilibrium $\s'$ that satisfies $EP_{V_i}(\s')\ge
T_i-\epsilon$ for $i=1, \dots, n$ and hence no Nash equilibrium $\s$
(not necessarily discrete) such that $EP_{V_i}(\s)\ge T_i$ for $i=1,
\dots, n$.

The running time analysis is similar to that for
Theorem~\ref{thm:bestnash}; we conclude that the running time of our
algorithm is $O(nN^3)=O(\max\{nP_{\max}^3/\epsilon^3,
n^4/\epsilon^3\})$.
\end{proof}

\begin{remark}
Theorem~\ref{thm:besteach} can be extended to trees of bounded degree 
in the same way as Theorem~\ref{cor:tree}.
%We should really give more detail.
\end{remark}

\subsection{Exact Computation}\label{sub:exact}
Another approach to finding Nash equilibria with bounded payoffs is
based on inductively computing the subsets of the best response
policies of all players so as to exclude the points that do not
provide sufficient payoffs to some of the players.  Formally, we say
that a strategy $v$ of the player $V$ is a {\em potential best
response to a strategy $w$ of its parent $W$ with respect to a
threshold vector $\T=(T_1, \dots, T_n)$}, (denoted by $v\in\pbr_{V}(w,
\T)$) if there is an equilibrium in the instance $G_{(V,W),W=w}$ in
which $V$ plays mixed strategy~$v$ and the payoff to any player $V_i$
downstream of $V$ (including $V$) is at least $T_i$.  The {\em best
response policy for $V$ with respect to a threshold vector $\T$} is
defined as $\PBR(W, V, \T)=\{(w, v)\mid v\in\pbr_V(w, \T), w\in[0,
1]\}$.

It is easy to see that if any of the sets $\PBR(V_{j}, V_{j-1}, \T)$,
$j=1, \dots, n$, is empty, it means that it is impossible to provide
all players with expected payoffs prescribed by $\T$. Otherwise, one
can apply the downstream pass of the original algorithm of~\cite{kls}
to find a Nash equilibrium. As we assume that~$V_n$ is a dummy vertex
whose payoff is identically $0$, the Nash equilibrium with these
payoffs exists as long as $T_n\le 0$ and $\PBR(V_{n}, V_{n-1}, \T)$ is
not empty.

Using the techniques developed in~\cite{egg}, it is not hard to show
that for any $j=1, \dots, n$, the set $\PBR(V_{j}, V_{j-1}, \T)$
consists of a finite number of rectangles, and moreover, 
one can compute $\PBR(V_{j+1}, V_{j}, \T)$ given
$\PBR(V_{j}, V_{j-1}, \T)$.  The advantage of this approach is that it
allows us to represent {\em all} Nash equilibria that provide required
payoffs to the players.  However, it is not likely to be practical,
since it turns out that the rectangles that appear in the
representation of $\PBR(V_{j}, V_{j-1}, \T)$ may have irrational
coordinates.

\begin{claim}
There exists a graphical game $G$ on a 3-vertex path $UVW$ and a
vector $\T=(T_1, T_2, T_3)$ such that $\PBR(V, W, \T)$ cannot be
represented as a union of a finite number of rectangles with rational
coordinates.
\end{claim}
\begin{proof}
We define the payoffs to the players in $G$ as follows.  The payoff to
$U$ is identically $0$, i.e., $P^0(U)=P^1(U)=0$.  Using
Claim~\ref{clm:any}, we select the payoffs to $V$ so that $P^0(V)=uw$,
$P^1(V)=P^0(V)+w-.8u-.1$, where $u$ and $w$ are the (mixed) strategies
of $U$ and $W$, respectively.  It follows that $V$ is indifferent
between playing 0 and 1 if and only if $w=f(u)=.8u+.1$; observe that
for any $u\in[0, 1]$ we have $f(u)\in[0, 1]$.  It is not hard to see
that we have
$$
\PBR(W, V)=[0, .1]{\times}\{0\}\cup[.1, .9]{\times}[0, 1]
\cup[.9, 1]{\times}\{1\}.
$$
The payoffs to $W$ are not important for our construction;
for example, set $P_0(W)=P_0(W)=0$.

Now, set $\T=(0, 1/8, 0)$, i.e., we are interested in Nash equilibria
in which $V$'s expected payoff is at least $1/8$.  Suppose $w\in[0,
1]$. The player $V$ can play a mixed strategy $v$ when $W$ is playing
$w$ as long as $U$ plays $u=f^{-1}(w)=5w/4-1/8$ (to ensure that $V$ is
indifferent between 0 and 1) and $P^0(V)=P^1(V)=uw=w(5w/4-1/8)\ge
1/8$. The latter condition is satisfied if $w\le (1-\sqrt{41})/20<0$
or $w\ge (1+\sqrt{41})/20$.  Note that we have
$.1<(1+\sqrt{41})/20<.9$.  For any other value of $w$, any strategy of
$U$ either makes $V$ prefer one of the pure strategies or does not
provide it with a sufficient expected payoff.  There are also some
values of $w$ for which $V$ can play a pure strategy (0 or 1) as a
potential best response to $W$ and guarantee itself an expected payoff
of at least $1/8$; it can be shown that these values of $w$ form a
finite number of segments in $[0, 1]$.  We conclude that any
representation of $\PBR(W, V, \T)$ as a union of a finite number of
rectangles must contain a rectangle of the form $[(1+\sqrt{41})/20,
w'']{\times}[v', v'']$ for some $w'', v', v''\in[0, 1]$.
\end{proof}

On the other hand, it can be shown 
that for any integer payoff matrices and threshold vectors and
any $j=1, \dots, n-1$, the sets $\PBR(V_{j+1}, V_j, \T)$ contain no
rectangles of the form $[u', u'']{\times}\{v\}$ or $\{v\}{\times}[w',
w'']$, where $v\in\reals\setminus\rationals$. This means that if
$\PBR(V_n, V_{n-1}, \T)$ is non-empty, i.e., there is a Nash
equilibrium with payoffs prescribed by $\T$, then the downstream pass
of the algorithm of~\cite{kls} can always pick a strategy profile that
forms a Nash equilibrium, provides a payoff of at least $T_i$ to the
player $V_i$, and has no irrational coordinates. Hence, unlike in the
case of the Nash equilibrium that maximizes the social welfare,
working with irrational numbers is not necessary, and the fact that
the algorithm discussed in this section has to do so can be seen as an
argument against using this approach.

\section{Other Criteria for Selecting a Nash Equilibrium}\label{fair}
In this section, we consider several other criteria that can be useful
in selecting a Nash equilibrium.

\subsection{Combining welfare maximization with\newline 
bounds on payoffs}
\label{combining}
In many real life scenarios, we want to maximize the social welfare
subject to certain restrictions on the payoffs to individual
players. For example, we may want to ensure that no player gets a
negative expected payoff, or that the expected payoff to player $i$ is
at least
%change "Delta" to "xi" so that we can use "Delta" for degree, which is
%more standard?
$P_{\max}^i-\xi$, where $P_{\max}^i$ is the maximum entry of $i$'s
payoff matrix and $\xi$ is a fixed parameter.  Formally, given a
graphical game $G$ and a vector $T_1, \dots, T_n$, let ${\mathcal S}$
be the set of all Nash equilibria $\s$ of $G$ that satisfy $T_i\le
EP_{V_i}(\s)$ for $i=1, \dots, n$, and let
$\hat{\s}=\argmax_{\s\in{\mathcal S}} EP(\s)$.

If the set ${\mathcal S}$ is non-empty, we can find a Nash equilibrium
$\hat{\s}'$ that is $\epsilon$-close to satisfying the payoff bounds
and is within $\epsilon$ from $\hat{\s}$ with respect to the total
payoff by combining the algorithms of Section~\ref{bestnash} and
Section~\ref{bounds}.

%Renamed to M^* because it is like M. Also, note new notation EP(s) etc
%(defined just before theorem3)
%nope: it is argmax, not max.
Namely, for a given $\epsilon>0$, choose $\delta$ as in the proof of
Theorem~\ref{thm:bestnash}, and let $X_i$ be the set of all discrete
strategies of player $V_i$ (for a formal definition, see the proof of
Theorem~\ref{thm:bestnash}).  Combining the proofs of
Theorem~\ref{thm:bestnash} and Theorem~\ref{thm:besteach}, we can see
that the strategy profile $\hat{\t}$ given by
$\hat{t}_i=\max\{x_{i}^j\mid x_i^j\le \hat{s}_i\}$ satisfies
$EP_{V_i}(\hat{\t})\ge T_i-\epsilon$,
$|EP(\hat{\s})-EP(\hat{\t})|\le\epsilon$.

Define $\hat{m}_i^{l, k}$ to be the maximum total payoff that 
$V_1, \dots, V_{i-1}$ can achieve if each $V_j$, $j\le i$, chooses a
strategy from $X_j$, for each $j<i$ the strategy of $V_j$ is a
potential best response to the strategy of $V_{j+1}$ and the payoff to
player $V_j$ is at least $T_j-\epsilon$, and, moreover, $V_{i-1}$
plays $x_{i-1}^l$, $V_i$ plays $x_i^k$.  If there is no way to choose
the strategies for $V_1, \dots, V_{i-1}$ to satisfy these conditions,
we set $m_i^{l, k}=-\infty$. The $\hat{m}_i^{l, k}$ can be computed by
dynamic programming similarly to the $m_i^{l, k}$ and $z_i^{l, k}$ in
the proofs of Theorems~\ref{thm:bestnash} and~\ref{thm:besteach}.
Finally, as in the proof of Theorem~\ref{thm:bestnash}, we use
$m_n^{l, k}$ to select the best discrete Nash equilibrium subject to
the payoff constraints.

Even more generally, we may want to maximize the total payoff to a
subset of players (who are assumed to be able to redistribute the
profits fairly among themselves) while guaranteeing certain expected
payoffs to (a subset of) the other players. This problem can be
handled similarly.

\subsection{A minimax approach}
\label{minimax}
A more egalitarian measure of the quality of a Nash equilibrium is the
minimal expected payoff to a player. The optimal solution with respect
to this measure is a Nash equilibrium in which the minimal expected
payoff to a player is maximal. To find an approximation to such a Nash
equilibrium, we can combine the algorithm of Section~\ref{bounds} with
binary search on the space of potential lower bounds. Note that the
expected payoff to any player $V_i$ given a strategy $\s$ always
satisfies $-P_{\max}\le EP_{V_i}(\s)\le P_{\max}$.

For a fixed $\epsilon>0$, we start by setting $T'=-P_{\max}$,
$T''=P_{\max}$, $T^*=(T'+T'')/2$.  We then run the algorithm of
Section~\ref{bounds} with $T_1=\dots=T_n=T^*$. If the algorithm
succeeds in finding a Nash equilibrium $\s'$ that satisfies
$EP_{V_i}(\s')\ge T^*-\epsilon$ for all $i=1, \dots, n$, we set
$T'=T^*$, $T^*=(T'+T'')/2$; otherwise, we set $T''=T^*$,
$T^*=(T'+T'')/2$ and loop.  We repeat this process until
$|T'-T''|\le\epsilon$.  It is not hard to check that for any
$p\in\reals$, if there is a Nash equilibrium $\s$ such that
$\min_{i=1, \dots, n}EP_{V_i}(\s)\ge p$, then our algorithm outputs a
Nash equilibrium $\s'$ that satisfies $\min_{i=1, \dots,
n}EP_{V_i}(\s)\ge p-2\epsilon$.  The running time of our algorithm is
$O(\max\{nP_{\max}^3\log\epsilon^{-1}/\epsilon^3,
n^4\log\epsilon^{-1}/\epsilon^3\})$.

\subsection{Equalizing the payoffs}
\label{equalising}
When the players' payoff matrices are not very different, it is
reasonable to demand that the expected payoffs to the players do not
differ by much either. We will now show that Nash equilibria in this
category can be approximated in polynomial time as well.

Indeed, observe that the algorithm of Section~\ref{bounds} can be
easily modified to deal with upper bounds on individual payoffs rather
than lower bounds. Moreover, we can efficiently compute an
approximation to a Nash equilibrium that satisfies both the upper
bound and the lower bound for each player.  More precisely, suppose
that we are given a graphical game $G$, $2n$ rational numbers $T_1,
\dots, T_n, T'_1, \dots, T'_n$ and $\epsilon>0$.  Then if there exists
a strategy profile $\s$ such that $\s$ is a Nash equilibrium for $G$
and $T_i\le EP_{V_i}(\s)\le T'_i$ for $i=1, \dots, n$, we can find a
strategy profile $\s'$ such that $\s'$ is a Nash equilibrium for $G$
and $T_i-\epsilon\le EP_{V_i}(\s')\le T_i'+\epsilon$ for $i=1, \dots,
n$.  The modified algorithm also runs in time\linebreak
$O(\max\{nP_{\max}^3/\epsilon^3, [4]n^4/\epsilon^3\})$.

This observation allows us to approximate Nash equilibria in which all
players' expected payoffs differ by at most $\xi$ for any fixed
$\xi>0$. Given an $\epsilon>0$, we set $T_1=\dots=T_n= -P_{\max}$,
$T'_1=\dots=T'_n= -P_{\max}+\xi+\epsilon$, and run the modified
version of the algorithm of Section~\ref{bounds}. If it fails to find
a solution, we increment all $T_i, T'_i$ by $\epsilon$ and loop. We
continue until the algorithm finds a solution, or $T_i\ge P_{\max}$.

Suppose that there exists a Nash equilibrium $\s$ that satisfies\linebreak
$|EP_{V_i}(\s)-EP_{V_j}(\s)|\le \xi$ for all $i, j=1, \dots, n$. Set
$r=\min_{i=1, \dots, n}EP_{V_i}(\s)$; we have $r\le EP_{V_i}(\s)\le
r+\xi$ for all $i=1, \dots, n$. There exists a $k\ge 0$ such that
$-P_{\max}+(k-1)\epsilon \le r\le -P_{\max}+k\epsilon$. During the
$k$th step of the algorithm, we set
$T_1=\dots=T_n=-P_{\max}+(k-1)\epsilon$, i.e., we have $r-\epsilon\le
T_i\le r$, $r+\xi\le T'_i\le r+\xi+\epsilon$. That is, the Nash
equilibrium $\s$ satisfies $T_i\le r\le EP_{V_i}(\s)\le r+\xi\le
T'_i$, which means that when $T_i$ is set to
$-P_{\max}+(k-1)\epsilon$, our algorithm is guaranteed to output a
Nash equilibrium $\t$ that satisfies $r-2\epsilon\le T_i-\epsilon\le
EP_{V_i}(\t)\le T'_i+\epsilon\le r+\xi+2\epsilon$.  We conclude that
whenever such a Nash equilibrium $\s$ exists, our algorithm outputs a
Nash equilibrium $\t$ that satisfies $|EP_{V_i}(\t)-EP_{V_j}(\t)|\le
\xi+4\epsilon$ for all $i, j=1, \dots, n$. The running time of this
algorithm is $O(\max\{nP_{\max}^3/\epsilon^4, n^4/\epsilon^4\})$.

Note also that we can find the smallest $\xi$ for which such a Nash
equilibrium exists by combining this algorithm with binary search over
the space $\xi=[0, 2P_{\max}]$. This identifies an approximation to
the ``fairest'' Nash equilibrium, i.e., one in which the players'
expected payoffs differ by the smallest possible amount.

Finally, note that all results in this section can be extended to
bounded-degree trees.

\section{Conclusions}\label{sec:conclusions}

We have studied the problem of equilibrium selection in graphical
games on bounded-degree trees. We considered several criteria for
selecting a Nash equilibrium, such as maximizing the social welfare,
ensuring a lower bound on the expected payoff of each player,
etc. First, we focused on the algebraic complexity of a social
welfare-maximizing Nash equilibrium, and proved strong negative
results for that problem. Namely, we showed that even for graphical
games on paths, any algebraic number $\alpha\in[0, 1]$ may be the only
strategy available to some player in all social welfare-maximizing
Nash equilibria. This is in sharp contrast with the fact that
graphical games on trees always possess a Nash equilibrium in which
all players' strategies are rational numbers.

We then provided approximation algorithms for selecting Nash
equilibria with special properties. While the problem of finding
approximate Nash equilibria for various classes of games has received
a lot of attention in recent years, most of the existing work aims to
find $\epsilon$-Nash equilibria that satisfy (or are $\epsilon$-close
to satisfying) certain properties.  Our approach is different in that
we insist on outputting an exact Nash equilibrium, which is
$\epsilon$-close to satisfying a given requirement. As argued in the
introduction, there are several reasons to prefer a solution that
constitutes an exact Nash equilibrium.

Our algorithms are fully polynomial time approximation schemes, i.e.,
their running time is polynomial in the inverse of the approximation
parameter $\epsilon$, though they may be pseudopolynomial with respect
to the input size. Under mild restrictions on the inputs, they can be
modified to be truly polynomial.  This is the strongest positive
result one can derive for a problem whose exact solutions may be hard
to represent, as is the case for many of the problems considered
here. While we prove most of our results for games on a path, they can be
generalized to any tree for which the best response policies have
compact representations as unions of rectangles. In the appendix, 
we show how to do this for the algorithm that finds a payoff-maximizing 
Nash equilibrium; other algorithms can be treated similarly.

Further work in this vein could include extensions to the
kinds of guarantees sought for Nash equilibria, such as guaranteeing
total payoffs for subsets of players, selecting equilibria in which
some players are receiving significantly higher payoffs than their 
peers, etc. At 
the moment however,
it is perhaps more important to investigate whether  Nash
equilibria of graphical games can be computed in a decentralized
manner, in contrast to the algorithms we have introduced here.

It is natural to ask if our results or those of~\cite{egg}
can be generalized to games with three or more actions. However, it
seems that this will make the analysis significantly more difficult.
In particular, note that one can view the bounded payoff games as a
very limited special case of games with three actions per player.
Namely, given a two-action game with payoff bounds, consider a game in
which each player $V_i$ has a third action that guarantees him a
payoff of $T_i$ no matter what everyone else does.  Then checking if
there is a Nash equilibrium in which none of the players assigns a
non-zero probability to his third action is equivalent to checking if
there exists a Nash equilibrium that satisfies the payoff bounds in
the original game, and Section~\ref{sub:exact} shows that finding an
exact solution to this problem requires new ideas.

Alternatively it may be interesting to look for similar results in the
context of correlated equilibria (CE), especially since the best CE
may have higher value (total expected payoff) than the best NE. The
ratio between these values is called the {\em mediation value}
in~\cite{ashlagi}.  It is known from~\cite{ashlagi} that the mediation
value of 2-player, 2-action games with non-negative payoffs is at most
$\frac{4}{3}$, and they exhibit a 3-player game for which it is
infinite. Furthermore, a 2-player, 3-action example from~\cite{ashlagi}
also has infinite mediation value.

\section{Appendix}

\noindent{\bf Theorem~\ref{cor:tree}\ }
Let $G$ be an $n$-player graphical game on a tree in 
which
each node has at most~$\Delta$ children. Suppose we are given a set of
best-response policies for~$G$ in which each best-response policy~$B(V,U)$ 
is
represented by a set of rectangles with at most $q(n)$ event points. For 
any
$\epsilon>0$, there is an algorithm that constructs a Nash 
equilibrium~$\s'$
for~$G$ that satisfies $EP(\s') \ge M(G) - \epsilon$. The running time of 
the
algorithm is polynomial in $n$, $P_{\mathrm{max}}$ and $\epsilon^{-1}$ 
provided
that the tree has bounded degree (that is, $\Delta=O(1)$) and $q(n)$ is
a polynomial in~$n$. In particular, if
$$N=\max((\Delta+1)q(n)+1,
n 2^{\Delta+2}(\Delta+2)P_{\max}\epsilon^{-1}
)$$ and $\Delta>1$
then the
running time is $O(n \Delta {(2N)}^\Delta$.
\medskip

\begin{proof}
Let  $\delta=1/N$ and let $X=\{j\delta \mid j=0, \dots, N\}$.
Consider the set of best-response policies for~$G$.
We say that a point $v\in[0,1]$ is an event point for a player~$V$
with parent~$W$ if either $v$ is a $V$-event point of $B(W,V)$ or, for 
some
child~$U$ of~$V$, $v$ is a
$V$-event point of $B(V,U)$. Let~$X'_V$ be the set of event points 
for~$V$.
Since $N\geq(\Delta+1)q(n)+1$, $|X'_V| \leq N-1$.
Let $X_V$ be a size-$2N$ superset of~$
X\cup X'_V$.

We will refer to the strategies in $X_V$ as {\em discrete} strategies
of player $V$. A strategy profile in which each player has a discrete
strategy will be referred to as a {\em discrete} strategy profile.
We will now show that
there is a discrete strategy profile~$\t$ which
is a Nash equilibrium and in which the total payoff is at least
$M(G)-\epsilon$.

Let $\s$ be a Nash equilibrium that maximizes social welfare
(so $EP(\s)=M(G)$).
For every player $V$, let $t_V
=\max\{x_{V}\in X_V\mid x_V\le s_V\}$.

First, we will show that the strategy profile $\t$ is a Nash equilibrium 
for $G$.
Consider a player~$V$ with parent~$W$.
Let $R=[w_1, w_2]{\times}[v_1, v_2]$ be the rectangle
in $\PBR(W, V)$ that contains $(s_W, s_V)$.
As $w_1$ is a $W$-event point of $\PBR(W, V)$,
we have $w_1\le t_W$ and $t_W\leq s_w \leq w_2$, so the point $(t_W, s_V)$
is inside $R$. Similarly,
the point $v_1$ is a $V$-event point
of $\PBR(W,V)$, so we have $v_1\le t_V$,
and therefore the point $(t_W, t_V)$ is inside $R$.
This means that for any player~$V$ with parent~$W$, we have
$t_{V}\in\pbr_{V}(t_W)$, which implies that
$\t$ is a Nash equilibrium for $G$.

Now, let us estimate the expected loss in social welfare
caused by playing $\t$ instead of $\s$.
\begin{lemma}\label{3deltatree}
For any pair of strategy profiles $\t, \s$ such that, for
all players~$V$, $|t_V-s_V|\le\delta$,
we also have
$$|EP_{V}(\s)-EP_{V}(\t)|\le 2^{\Delta+2}(\Delta+2)P_{\max}\delta$$
for all~$V$.
\end{lemma}
\begin{proof}

Fix any player~$U_1$.
Let~$U_0$ be his parent, and let $U_2,\ldots,U_d$ be his children.
For $\sigma\in \{0,1\}^{d+1}$, let
$P_{\sigma}$ be the payoff to~$U_1$ when, for $\ell\in\{0,\ldots,d+1\}$,
$U_\ell$ plays $\sigma_\ell$.
Let $t_{\sigma}$ be the probability that this event occurs
according to strategy profile~$\t$.
Formally, let
$$ \Psi(z,b) = \Big\{
\begin{array}{cc}
z, & \mbox{if $b=1$},\\
1-z, & \mbox{if $b=0$.}\\
\end{array}
$$
Then
$$t_{\sigma} = \prod_{\ell=0}^{d+1} \Psi(t_{U_\ell},\sigma_\ell).$$
Similarly, let $s_{\sigma}$ be the probability that this event occurs
according to strategy profile~$\s$.
Then
\begin{align*}
|EP_{V}(\s)-EP_{V}(\t)|
&= \left|\sum_\sigma P_\sigma t_\sigma - \sum_\sigma P_\sigma s_\sigma 
\right|\\
&\le \sum_{\sigma}|P_{\sigma}(t_{\sigma}-s_{\sigma})|\\
&\le
2^{\Delta+2} P_{\max}\max_{\sigma}|t_{\sigma}-s_{\sigma}|.\\
\end{align*}
We will now show that, for any $\sigma$,
$|t_{\sigma}-s_{\sigma}| \leq (\Delta+2)\delta$, which implies the lemma.
For $r\in\{0,\ldots,d+2\}$, let
$$z_{r,\sigma} = \prod_{\ell=0}^{r-1} \Psi(t_{U_\ell},\sigma_\ell)
\prod_{\ell=r}^{d+1} \Psi(s_{U_\ell},\sigma_\ell).$$

Since $|t_{U_\ell} - s_{U_\ell}|\leq \delta$, we have, for any 
$j\in\{0,1\}$,
$|\Psi(t_{U_\ell},j)-\Psi(s_{U_\ell},j)|\leq \delta$.
So $|z_{\ell,\sigma}-z_{\ell-1,\sigma}|\leq \delta$.
Then
\begin{align*}
|t_{\sigma}-s_{\sigma}| &= |z_{d+2,\sigma}-z_{0,\sigma}|
= \left|
\sum_{\ell=1}^{d+2} \left(z_{\ell,\sigma}-z_{\ell-1,\sigma}\right)
\right|\\
&\leq \sum_{\ell=1}^{d+2} |z_{\ell,\sigma}-z_{\ell-1,\sigma}|
\leq  (\Delta+2)\delta.\\
\end{align*}\end{proof}

So by Lemma~\ref{3deltatree},
the expected loss in social welfare
caused by playing $\t$ instead of $\s$
is at most
$n 2^{\Delta+2}(\Delta+2)P_{\max}\delta$.
Since $N=1/\delta$ is at least
$n 2^{\Delta+2}(\Delta+2)P_{\max}\epsilon^{-1}$, this is at 
most~$\epsilon$.

We will now show that how to find the best discrete
Nash equilibrium (with respect to the social welfare)
using dynamic programming.
As $\t$ is a discrete strategy profile,
the strategy profile found by our algorithm
will be at least as good as $\t$.

For a player~$V$ with $d$ children and a string
$\tau\in\{1,\ldots,2N\}^{d+1}$,
define $m_{V,\tau}$ to be the maximum total payoff that proper descendants 
of~$V$
can achieve in a strategy profile satisfying the following conditions,
where $x_V$ denotes the $\tau_0$'th strategy from the discrete strategy 
set~$X_V$
of player~$V$, and for every $\ell\in\{1,\ldots,d\}$,
$U_\ell$ is the $\ell$'th child of~$V$ and $x_{U_\ell}$ is the
$\tau_\ell$'th strategy from the discrete strategy set~$X_{U_\ell}$.
\begin{itemize}
\item
$V$ plays $x_{V}$.
\item For every
$\ell\in\{1,\ldots,d\}$, $U_\ell$ plays $x_{U_\ell}$.
\item All descendants of~$V$ play strategies from their discrete strategy 
sets.
\item For every proper descendant~$U$ of~$V$,
the strategy chosen by~$U$ is a potential best response to the strategy 
chosen by its parent.
\end{itemize}
If no strategy profile satisfies these conditions, $m_{V,\tau}=-\infty$.

For any leaf~$V$ of the tree and any string~$\tau$, $m_{V,\tau}=0$.
Here is how the algorithm computes $m_{V,\tau}$ for an internal
vertex~$V$, assuming it
has already computed $m_{U,\tau'}$ for all proper descendants
$U$ of~$V$ and all strings~$\tau'$.
First, it checks every child $U_\ell$ of~$V$, to see whether
$x_{U_\ell} \in \pbr_{U_\ell}(x_V)$. If this is not the case, the 
algorithm
sets $m_{V,\tau}=-\infty$ and finishes.
Now let $r_\ell$ denote the number of children of~$U_\ell$ (which may be 
zero)
and let ${\widehat{m}}_{\ell}$ equal
$$
\max \left\{
m_{U_\ell,\tau'}+P_{\ell,\tau'}
\mid
\tau'\in{\{1,\ldots,2N\}}^{r_\ell+1},
{\tau'}_0=\tau_\ell
\right\},$$
where $P_{\ell,\tau'}$ denotes the payoff to~$U_\ell$ when $V$ plays~$x_V$
and $U_\ell$ and its children play according to~$\tau'$
(the $i$th child of $U_\ell$ plays the $\tau'_i$th element of its
discrete strategy set and $U_\ell$ plays $x_{U_\ell}$).
Then
$$m_{V,\tau} = \sum_{\ell=1}^d {\widehat{m}}_{\ell},$$
and this can be computed in polynomial time by considering
each possible~$\tau'$
(at most ${(2N)}^{\Delta}$ of them)
for each child~$U_\ell$ (of which there are at most~$\Delta$).

Finally, suppose that we have computed all of the values $m_{V,\tau}$.
We assumed without loss of generality
(see Section~\ref{defs}) that the
root of the tree is a node~$V$ which has constant payoff~$0$, 
independently
of the action chosen by its singleton child~$U_1$.
To find the discrete strategy that maximizes social welfare, we just
choose the $\tau\in\{1,\ldots,2N\}^2$ that
maximises $m_{V,\tau}$. This value, $m_{V,\tau}$ is the
social welfare achieved by the algorithm.
The discrete strategy that achieves this social welfare can now be
constructed by standard dynamic programming techniques.
\end{proof}
\end{document}